\documentclass[12pt, twoside]{article}
\usepackage{a4wide,amssymb,cite}
\usepackage{epsfig,axodraw}

\def\de{\partial}
\def\a{\alpha}
\def\b{\beta}
\def\g{\gamma}
\def\G{\Gamma}
\def\d{\delta}

\def\e{\eta}
\def\la{\lambda}
\def\La{\Lambda}
\def\k{\kappa}
\def\m{\mu}
\def\n{\nu}
\def\r{\rho}
\def\o{\omega}
\def\s{\sigma}
\def\S{\Sigma}

\def\p{\pi}
{\rm }
\def\F{\Phi}
\def\vf{\varphi}

\def\th{\theta}
\def\de{\partial}

\def\x{\chi}

\newcommand{\be}{\begin{equation}}
\newcommand{\ee}{\end{equation}}
\newcommand{\bea}{\begin{eqnarray}}
\newcommand{\eea}{\end{eqnarray}}
\newcommand{\beqar}{\begin{eqnarray*}}
\newcommand{\eeqar}{\end{eqnarray*}}
\newcommand{\eg}{{\it e.g.,}\ }
\newcommand{\ie}{{\it i.e.,}\ }
\newcommand{\reef}[1]{(\ref{#1})}
\newcommand{\nn}{\nonumber}

\begin{document}

\begin{titlepage}

\rightline{hep-th/0611311}
\rightline{November 2006}

\begin{centering}
\vspace{1cm}
{\Large {\bf Regularization of conical singularities\\ in warped six-dimensional compactifications}}\\

\vspace{1cm}

 {\bf Eleftherios Papantonopoulos}$^{a}$, {\bf Antonios~Papazoglou} $^{b,c,d}$ \\ and  {\bf Vassilios Zamarias}$^{a}$ \\
\vspace{.2in}

$^{a}$ Department of Physics, National Technical University of
Athens,\\
Zografou Campus GR 157 73, Athens, Greece \\
\vspace{2mm}
$^{b}$ APC\footnote{UMR 7164 (CNRS, Universit\'e Paris 7, CEA, Observatoire de Paris)}, 11 place Marcelin Berthelot,\\
F 75005 Paris Cedex 05, France\\
\vspace{2mm}
$^{c}$ GReCO/IAP\footnote{UMR 7095 (CNRS, Universit\'e Paris 6)}, 98 bis Boulevard Arago,\\
F 75014 Paris, France\\
\vspace{2mm}
$^{d}$ \'Ecole Polytechnique F\'ed\'erale de Lausanne,\\ Institute of Theoretical Physics,\\
SB ITP LPPC BSP 720, CH 1015, Lausanne, Switzerland.

\end{centering}
\vspace{1cm}

\begin{abstract}

We study the regularization of the codimension-2 singularities in
six-dimensional  Einstein-Maxwell axisymmetric models with
warping.  These singularities are replaced by codimension-1
branes of a ring form, situated around the axis of symmetry. We assume that 
there is a brane scalar field with Goldstone dynamics, which is known to  
generate a brane energy momentum tensor of a particular structure necessary for  the  above regularization to be successful.  We
study these compactifications in both a non-supersymmetric and
a supersymmetric setting. We see that in the non-supersymmetric
case, there is a restriction to the admissible warpings and
furthermore to the quantum numbers of the bulk gauge field and the
brane scalar field. On the contrary, in the supersymmetric case,
the warping can be arbitrary.

\end{abstract}

\vspace{1cm}
\begin{flushleft}

 E-mail addresses: lpapa@central.ntua.gr,  papazogl@iap.fr, zamarias@central.ntua.gr

\end{flushleft}
\end{titlepage}


\section{Introduction}

There has been considerable attention the past years at brane
models  with codimension-2 branes in six-dimensional theories.
This attention has mostly to do with an
interesting property of their vacuum energy and has inspired the
construction of models which may ameliorate the cosmological
constant problem. The latter property is the fact that their
vacuum energy instead of curving their world volume, merely
changes the deficit angle in the geometry of the surrounding bulk
\cite{luty}. Thus in principle,  models which decouple the
curvature of the brane from the brane vacuum energy can be constructed
 (selftuning models).

The most thoroughly discussed models of this kind  were the ones
with  a bulk gauge field coupled to gravity \cite{selftune} (for
another possibility see \cite{sigma,swirl}). These models have
generally a monopole solution which spontaneously compactifies the
internal two-dimensional space \cite{spontaneous}. This internal
space can in principle have some conical defects which support
codimension-2 branes. A usual assumption which we adopt also in
the present paper is axial symmetry (see \cite{ll} for a more general case). This restricts the number of
the codimension-2 branes to be at most two, situated at the north
and south pole of the internal space.  The size of spontaneously
compactified internal space is classically determined in the case
of a  purely Einstein-Maxwell system \cite{spontaneous,jap}, but
behaves as  a modulus in its supersymmetrized version
\cite{SS,Gibbons,burgess2}, where a dilaton with appropriate
coupling to the other fields is also present (for the stability of
these models see \cite{japstab,stability}). The selftuning
property of these models is ruined however, by the flux
quantization condition \cite{fine-tuning}, unless one admits
solutions with singularities more severe than conical
\cite{non-conical}.

The study of gravity on the codimension-2 branes is a difficult
issue.  Any simplistic way to discuss non-trivial brane geometries
results to bulk singularities at the position of the brane much
more serious than conical \cite{ClineGiova}. One way to confront
this problem is to complicate the gravity dynamics by adding for
example a Gauss-Bonnet term in the bulk or an induced gravity term
on the brane \cite{GB}. This approach, however, leads to rather
restrictive constraints on the matter content of the brane
\cite{GBconstr}. The most natural procedure is to consider a
regularized version of the codimension-2 brane in Einstein
gravity, where the brane acquires some thickness \cite{thick} in
its transverse directions.

The latter approach has been followed in \cite{PST} in its
simplest form, where the codimension-2 branes were replaced by
thin ring-like codimension-1 branes warped around the axis of
symmetry. The space close to the conical tip, which is cut by the
codimension-1 brane, is replaced by appropriate bulk caps. This
kind of regularization requires a specific form of the brane
energy momentum tensor to work, which can be provided by a brane
scalar field  with  Goldstone-like  derivative couplings to the
bulk gauge field. This Goldstone feature can be natural if this
scalar field originates from a Higgs field whose radial component
has been integrated out \cite{burgess1} at low enough energies. In
\cite{PST} the regularization of the unwarped non-supersymmetric
model had been provided and gravity on the codimension-1 brane was
discussed.

In the present paper we generalize the above procedure by
considering this type of regularization for codimension-2 brane
models with general warping. Furthermore, we repeat the
regularization also for a gauged  supergravity model. We find that
this regularization in the non-supersymmetric case cannot work for
general warping (or equivalently for arbitrary monopole and brane
scalar field quantum numbers). On the contrary, in the
supersymmetric case no such constraint is present. In the next two
sections we will present the solutions with the codimension-2
branes and then the regularization approach followed by the
solutions with codimension-1 branes. At the end we will conclude
and suggest a possible application of these regularized
compactifications.

\section{Non-supersymmetric warped compactifications}

We discuss first the six-dimensional Einstein-Maxwell system which
was originally found to spontaneously compactify an internal
two-dimensional space \cite{spontaneous}. We suppose that we have
a six-dimensional gauge field ${\cal A}_M$ (with field strength
${\cal F}_{MN}$) coupled to gravity in the presence of a bulk
cosmological constant $\La_0$. The general axisymmetric solutions
of such a model, involve compact spaces with sphere topology and
two codimension-2 singularities at the poles of the deformed
sphere \cite{jap}.  The dynamics of this system can be encoded in
the  following action \be S= \int d^6x \sqrt{-g} \left( {M^4 \over
2} R - \La_0 -{1 \over 4}{\cal F}^2 \right) - T_\pm\int d^4x
\sqrt{-\g_{\pm}}~, \ee where brane actions are added to the bulk
part, with brane tensions  $T_\pm$ for the codimension-2 branes
situated at the north and south pole respectively. The tensor
$\g^{\pm}_{\m\n}$ is the induced metric on the two 3-branes. The
quantity $M$ is the six-dimensional Planck mass.

The equations of motion in the bulk are easily obtained by
variation of the metric and the gauge field and read \bea
R_{MN}&=& {1 \over  M^4}\left[{\La_0 \over 2}g_{MN} +{\cal F}_{MK}{\cal F}^{~K}_N - {1 \over 8}{\cal F}^2 g_{MN}\right] ~,   \label{Einstein}\\
\de_M \left( \sqrt{-g}{\cal F}^{MN}\right)&=&0 ~. \label{Gauge}
\eea

In the following, we will first discuss the background solution
with codimension-2 branes and then we will present the way to
regularize these branes by lowering their codimension.

\subsection{The general solution with codimension-2 branes}  \label{co2back}

Let us recall the general bulk axisymmetric solution of the above
system,  presented in \cite{jap}. The solution can be obtained
after a double Wick rotation from the one for the six-dimensional
Reissner-Nordstr\"om black hole. Thus, the metric and the only
non-zero component of the gauge field strength are \bea
ds_6^2&=& \r^2 \eta_{\m\n}dx^\m dx^\n  + {d\r^2 \over F} + c_0^2 F d\th^2 ~,\label{metricbhmis}\\
{\cal F}_{\r\th}&=& -{b_0 \over \r^4}~, \label{metricbh} \\
{\rm with} ~~~ M^4 F(\r)&=&-{\La_0 \over 10}\r^2 -{b_0^2 \over 12
c_0^2 \r^6}+{\m_0 \over \r^3}~, \label{gaugebh} \eea where $c_0$
is a constant which can be absorbed in the angular coordinate.
The reason why we keep it in the metric will be explained later.
Note that the dimensions of the various quantities in the solution
are $[\La_0]=[b_0^2]=[\m_0]=[F^3]=M^6$, $[\r]=[c_0]=M^0$, and
$[\th]=M^{-2}$.

The metric function $F(\r)$ has generically many real roots.
Without loss of generality, we will suppose that  two of these
roots are positive and we will consider the space
$0<\r_-<\r<\r_+$. From the condition that $F(\r_\pm)=0$, we can
express the parameters $b_0$ and $\m_0$ as a function of $\r_+$
and $\a \equiv \r_- / \r_+$ [$0 < \a \leq 1$] as \bea
b_0^2&=&{6 \over 5} c_0^2 \La_0~ \r_+^8 \a^3 {1- \a^5 \over 1-\a^3} \label{bexpr}~,\\
\m_0&=&{\La_0 \over 10}\r_+^5 {1- \a^8 \over 1-\a^3}~.
\eea

The parameter $\a$ expresses the degree of warping of the
four-dimensional part of the metric as it will be evident shortly.
The unwarped case corresponds to the value $\a=1$, when the two
roots of the function $F$ coincide. The latter limit is singular
in the present gauge, and thus it would be difficult to compare
results with the unwarped case. For this reason, we follow the
prescription of \cite{japstab} and we make the coordinate
transformation \bea
\r=\r_+ z ~~ {\rm with} ~~ z(r)={1 \over 2}[(1-\a)r+(1+\a)] ~, \\
\vf= {\La_0 \over M^4}\r_+(1-\a) \th ~.
\eea

Then in the new coordinates the solution
\reef{metricbhmis}-\reef{gaugebh} becomes \bea
ds_6^2&=&  z^2 \r_+^2 \eta_{\m\n} dx^\m dx^\n + R_0^2\left[ {dr^2 \over f} +c_0^2 f ~d\vf^2 \right]~,\\
{\cal F}_{r \vf}&=& - c_0 R_0 M^2 S\cdot {1 \over z^4}~,\\
{\rm with} ~~~ f&=&  {4  R_0^2 \over \r_+^2(1-\a)^2}F ~,\\
{\rm and}~~~S(\a)&=& \sqrt{{3 \over 5}\a^3{1-\a^5 \over 1-\a^3}}~,
\eea with  $R_0^2=M^4/(2\La_0)$ a quantity representing the
average radius of the internal space. In the limit $\a \to 1$, we
have that $S\to 1$. Note that $[f]=[\vf]=M^0$. From the above
definition we can verify that $f(r)$ depends only on the parameter
$\a$ as \be f(r)= {1 \over 5 (1-\a)^2}\left[ -z^2 + {1-\a^8 \over
1-\a^3}\cdot{1 \over z^3 }-\a^3{1-\a^5 \over 1-\a^3}\cdot{1 \over
z^6} \right]~. \ee

The range of the angular coordinates $\th$ and $\vf$ have not been
specified yet. We can in general take $\vf \in [0,2\pi \xi)$. The
parameter $\xi$ plays the same role as the parameter $c_0$ and by
a coordinate transformation we can keep one of them and set the
other to unity.  In other words, only the product of $\xi$ and $c_0$ 
has physical meaning.  We will keep however both of them in the following
with the purpose of making an easy comparison with the unwarped
case, choosing them appropriately.

The two codimension-2 branes are situated at $\r=\r_\pm$, or in
the new  coordinates at $r(\r_\pm)=\pm 1$ with  $z(\r_+)=1$,
$z(\r_-)=\a$. Expanding the function $f$ at $r \to \pm 1$ we get
$f \to 2 (1\mp r) X_\pm$ with the quantities $X_\pm$ given by \bea
X_+&=& {5+3\a^8-8\a^3 \over 20 (1-\a)(1-\a^3)} \label{X+} ~,\\
X_-&=& {3+5\a^8-8\a^5 \over 20 \a^4(1-\a)(1-\a^3)} ~.\label{X-}
\eea

Then the metric around  $r=\pm 1$ reads
\be
ds_6^2 \approx z^2(\pm 1) \r_+^2 \eta_{\m\n} dx^\m dx^\n + {R_0^2 \over X_\pm} \left[ {dr^2 \over 2 (1\mp r)} + c_0^2 X_\pm^2  2 (1\mp r) d \vf^2 \right] \label{localconical}~.
\ee

The deficit angles at the two singularities are thus $\d_\pm=2\pi
(1-\b_\pm)$ with $\b_\pm=c_0 X_\pm \xi$. These singularities are
supported by codimension-2 branes with tensions related to the
deficit angles as $T_{\pm}=M^4 \d_\pm$. Note that the quantities
$\xi$ and $c_0$ appear correctly together as stressed before.

In the limit $\a \to 1$, the warping of the space disappears and
we have the case of a sphere with a deficit angle $\d=2\pi (1-c_0
\xi)$, since then  $X_+=X_-=1$. In this case, the exact metric is
rather simple because $f=(1-r^2)$. Thus we get \bea
ds_6^2 &=& \r_+^2 \eta_{\m\n} dx^\m dx^\n + R_0^2 \left[ {dr^2 \over  (1 - r^2)} + c_0^2  (1 - r^2) d \vf^2 \right]~,\\
{\cal F}_{r \vf}&=& -c_0 R_0 M^2~. \eea As can be easily checked,
the above metric coincides close to
 the branes with \reef{localconical} for $\a \to 1$. With the coordinate
  transformation $r=\cos \o$ we can obtain the more familiar form of the solution as in \cite{selftune}
\bea
ds_6^2 &=& \r_+^2 \eta_{\m\n} dx^\m dx^\n + R_0^2 \left[ d\o^2 + c_0^2  \sin^2 \o d \vf^2 \right]~,\\
{\cal F}_{\o \vf}&=& c_0 R_0 M^2 \sin\o ~.
\eea

At this point we will partially fix $\xi$ by postulating that it
is a function of $\a$ and that in the limit of $\a \to 1$ it is
$\xi(\a) \to 1$. With this limiting behaviour the deficit angle in
the unwarped case is given by $\d=2\pi (1-c_0)$. Thus, the
quantity $c_0$ appearing in warped metric \reef{metricbhmis}
corresponds to the deficit angle of the unwarped solution. But as
we allow warping, the quantity $\xi(\a)$ is also in principle
present in the range of the angular coordinate. Particular choices
of $\xi(\a)$ (always with the above-mentioned limit) can make
certain calculations easier.

The parameter $\r_+$ can be absorbed to a $x^\m$-coordinate
redefinition, so from now on we take $\r_+=1$. The flux
quantization of the above gauge field, gives us a quantization
condition for the deficit angle. To find the quantization
condition, we need the solution for the gauge field in two patches
of the manifold (north and south) which correctly reproduce the
flux when Stokes' theorem is applied. The gauge field in these
patches should vanish at the poles since we have assumed that the
branes carry no charge. These gauge field are given by \bea
{\cal A}^+_\vf&=&  c_0 R_0 M^2 \cdot {2 S \over 3(1-\a)}\left({1 \over z^3}-1 \right)\label{Anorth}~,\\
{\cal A}^-_\vf&=&  c_0 R_0 M^2 \cdot {2 S \over 3(1-\a)}\left({1 \over z^3}-{1 \over \a^3} \right)\label{Asouth}~,
\eea
which in the unwarped case limit $\a \to 1$  gives
\be
{\cal A}^\pm_\vf =  c_0 R_0 M^2 (-r\pm 1)~. \label{gaugeunwarped}
\ee

Then single valuedness of the gauge transformation at the
overlapping region, gives the following quantization condition
\bea
2 c_0 R_0 M^2  \xi e~Y &=& N   ~~~,~~~N \in \mathbb{Z}~,  \label{N} \\
{\rm with}~~~Y&=&{(1-\a^3) \over 3 \a^3 (1-\a)}~S~,
\eea
where $e$ is a unit fundamental charge. In the limit when $\a \to 1$, it is $Y \to 1$ as expected.

We can choose the function $\xi(\a)$ to simplify the expression of
the quantization condition, or the one of the quantities $\b_\pm$.
A choice $\xi=1/Y$ is particularly helpful. In this case the
quantization condition remains the same as in the case of no
warping and the deficit angles are given by $\b_\pm=c_0 X_\pm/Y$.

\subsection{The regularization}

\begin{figure}[t]
\begin{center}
\begin{picture}(200,160)(-60,-20)

\Text(45,150)[c]{$r=+1$}
\Text(45,-5)[c]{$r=-1$}

\Text(125,70)[c]{Bulk: $c_0,~R_0,~\a$}

\Text(125,130)[c]{Cap: $c_+,~R_+,~\a$}

\Text(125,5)[c]{Cap: $c_-,~R_-,~\a$}

\Text(-65,142)[c]{Brane:}
\Text(-65,129)[c]{$\la_+,~v_+$}
\Text(-65,114)[c]{$r=r_+$}

\Text(-65,20)[c]{Brane:}
\Text(-65,6)[c]{$\la_-,~v_-$}
\Text(-65,-10)[c]{$r=r_-$}

\Oval(-65,6)(22,38)(0)

\Oval(-65,128)(22,38)(0)

\LongArrow(-20,128)(25,128)
\LongArrow(-20,6)(33,6)

\epsfig{file=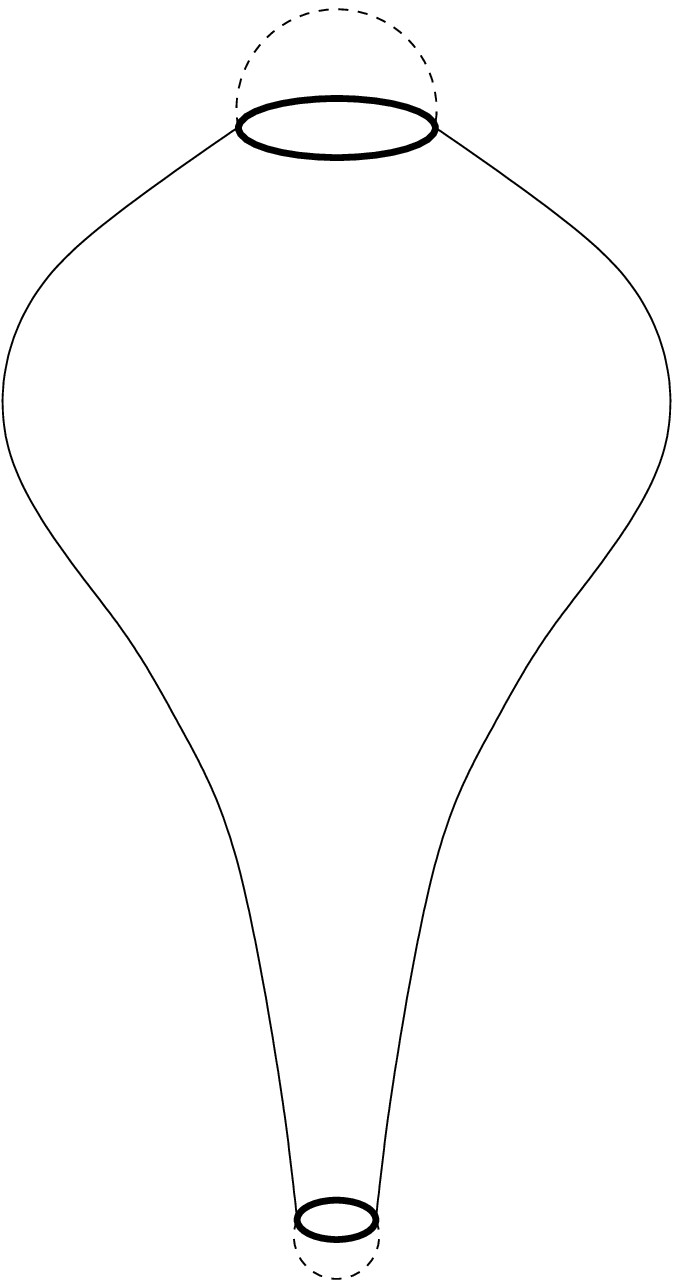,width=3cm,height=5cm}

\end{picture}
\caption{ The internal space where the codimension-2 singularities
have been  regularized with the introduction of ring-like
codimension-1 branes. As the positions of the rings tend to the
poles \ie when $r_\pm \to \pm 1$, we recover the conical brane
limit. The parameters of the action and the solution are denoted
in the appropriate part of the internal space. }
\label{internalfig}
\end{center}
\end{figure}

We would like to regularize this model by substituting  the
codimension-2 branes by branes of lower codimension. This is
possible, if we suppose that the space close to the  codimension-2
singularities is cut at $r=r_\pm$ by codimension-1 branes and is
then replaced by smooth caps, as done in \cite{PST} for the
unwarped case. Let us denote by ${\cal M}_0$ the bulk between the
codimension-1 branes, ${\cal M}_\pm$ the two caps and $\S_\pm$ the
two codimension-1 branes. The regularized space is drawn in
Fig.~\ref{internalfig}. Then the action of the system is written
as 
\be 
S= \int_{{\cal M}_i} d^6x \sqrt{-g} \left( {M^4 \over 2} R
- \La_i -{1 \over 4}{\cal F}^2 \right) - \int_{\S_\pm} d^5x
\sqrt{-\g_\pm}\left(\la_\pm +{v_\pm^2 \over 2}
(\tilde{D}_{\hat{\m}} \s_\pm)^2 +M^4 \{{\cal K}\}_\pm\right)~,
\ee 
 where $\g^\pm_{\hat{\m}\hat{\n}}=g_{MN}\de_{\hat{\m}} X^M \de_{\hat{\n}} X^N $ is the induced metric on the branes, with
$\hat{\m}=\{\m,\vf\}$ a five dimensional index and  $X^M$ the bulk coordinates of the brane. [Here, for the static brane  $X^N=(x^{\hat{\n}},r_\pm)$]. Also,
 $\La_i$, $i=0,\pm$ are the bulk cosmological constants in the three bulk
regions and  $\la_\pm$ are the tensions  of the codimension-1
branes. Furthermore, the branes have Goldstone-like scalar fields
$\s_\pm$ coupled to the bulk gauge field, through the combination
$\tilde{D}_{\hat{\m}} \s_\pm = \de_{\hat{\m}} \s_\pm -E a_{\hat{\m}}$, where $a_{\hat{\m}}={\cal A}_M \de_{\hat{\m}} X^M$ is the induced gauge field. Note that this in {\it not} the 
covariant derivative of $\s_\pm$ with respect to $a_{\hat{\m}}$. [Since we will consider static branes, we have simply $a_{\hat{\m}}={\cal A}_{\hat{\m}}$.] 
 The last term in the brane action is the Gibbons-Hawking term for each
brane. We denote $\{{\cal K}\}={\cal K}^{in} + {\cal K}^{out}$ the sum of the extrinsic
curvatures from each side of each brane. The extrinsic curvature is constructed
 using the normal to the brane $n_M$ which points {\it inwards to the corresponding part of the bulk}
each time (we use the conventions of \cite {ChamblinReal}). For more details see the Appendix B.

The scalar fields $\s_\pm$ originate from the phases of Higgs
scalar fields,  whose radial components have been integrated out
for low enough energies. If $C$ is the normalization of the Higgs
phase $e^{i C \s_\pm}$, we cannot directly read it from our
effective low energy  brane action. We can fix this ambiguity by
choosing, without loss of generality, $C=1/\xi$ where $\xi$ has to
do with the range of the angular coordinate $\vf$. This choice
simplifies the solution for the scalar fields $\s_\pm$.  As we
present in detail in Appendix A, the ``charge'' $E$ is related to
the charge of the parent Higgs field through $\xi$. If the Higgs
field has unit fundamental charge $e$, the ``charge'' $E$ is
related to the latter  as $E=\xi e$. It is important to note that
the ``charge'' $E$  need not be an integer multiple of the
fundamental charge $e$. This does not contradict the well known
fact that charges in compact spaces are quantized, since $E$ is
not a charge in the strict sense but just a parameter appearing in
the effective brane action.

The equations of motion in the bulk are that same as
\reef{Einstein}, \reef{Gauge},  where now we substitute $\La_i$
for the three bulk regions instead of $\La_0$.  The junction
conditions on the branes are obtained by matching the surface
terms from each side of the brane with the variation of the brane
action and read 
 \bea
\{\hat{K}_{\hat{\m}\hat{\n}}\}_\pm&=&-{1 \over M^4}\left[-\la_\pm \g_{\hat{\m}\hat{\n}}+ v_\pm^2 \left( \tilde{D}_{\hat{\m}} \s_\pm \tilde{D}_{\hat{\n}} \s_\pm -{1 \over 2} ( \tilde{D}_{\hat{\k}} \s_\pm)^2 \g_{\hat{\m}\hat{\n}} \right)\right]~, \label{Kjunction} \\
\{n_{M} {\cal F}^{M}_{~N}\de_{\hat{\k}} X^N  \}_\pm&=&-E v_\pm^2
\tilde{D}_{\hat{\k}} \s_\pm~, \label{Fjunction}
\eea 
where
$\hat{K}_{\hat{\m}\hat{\n}}=K_{\hat{\m}\hat{\n}}-K
\g_{\hat{\m}\hat{\n}}$, with $K_{\hat{\m}\hat{\n}}={\cal K}_{MN} \de_{\hat{\m}} X^M \de_{\hat{\n}} X^N$.  Finally, the equation of motion of the
brane  Goldstone fields is \be \de^{\hat{\m}} \tilde{D}_{\hat{\m}}
\s_\pm =0 ~. \ee

Before presenting the regularized solution, let us make a comment
regarding the implementation of the above junction conditions. The
metric in each side of the brane can be generally written as \be
ds_6^2=g_{\m\n}d x^\m d x^\n + g_{rr}dr^2 + g_{\vf\vf}d\vf^2~, \ee
with $g_{\m\n}$, $g_{\vf\vf}$ continuous functions as they cross
the brane. The derivatives of the latter metric components  are of
course discontinuous, which gives rise to the junction conditions.
However, the metric components $g_{r r}$ {\it need not} be
continuous. We can always make a radial coordinate redefinition $r
\to l(r)$ as in \cite{PST} \be g_{ll} dl^2 = [g^{out}_{r r}
\th(r_c-r)+ g^{in}_{r r} \th(r-r_c) ]dr^2~, \ee with $r_c$ the
brane position, where $g_{ll}$ is a continuous line element across
the brane. The above junctions are usually understood for such a
coordinate system. Then the normals on each side are opposite
vectors $n^{in}_M=-n^{out}_M$. Nevertheless, the above equations
still make sense for the case where  $g^{in}_{rr}\neq
g^{out}_{rr}$ at the position of the brane. In this case, the
``normals'' $n_M$ in each direction are {\it not} opposite vectors
(since they are normalized with different $g_{rr}$).

\subsection{The general solution with codimension-1 branes}

We can now present the background solution of the above
regularized system.  At the end, we will verify that taking the
limit of $r_\pm \to \pm 1$, the solution of Section \ref{co2back}
with the codimension-2 branes is correctly reproduced.

The solution in the bulk region ${\cal M}_0$ between the two
branes ($r_-<r<r_+$), is the one described in Section
\ref{co2back} \bea
ds_6^2&=&  z^2 \eta_{\m\n} dx^\m dx^\n + R_0^2\left[ {dr^2 \over f} +c_0^2 f~ d\vf^2 \right]~,\\
{\cal F}_{r \vf}&=&  -c_0 R_0 M^2 S \cdot {1 \over z^4}~, \eea
with $R_0^2=M^4/(2\La_0)$.  On the other hand, the caps ${\cal
M}_\pm$,  with $r_+<r<1$ and $-1<r<r_-$ respectively,  are
described by the solutions \bea
ds_6^2 &=&  z^2 \eta_{\m\n} dx^\m dx^\n + R_\pm^2\left[ {dr^2 \over f} +c_\pm^2 f d\vf^2 \right]~,\\
{\cal F}_{r \vf}&=&  -c_\pm R_\pm M^2 S \cdot {1 \over z^4}~, \eea
with $R_\pm^2=M^4/(2\La_\pm)$. Since the aim of the regularization
is not to have codimension-2 branes present, one should demand
that there is no deficit angle in the caps, \ie that  $\b_\pm=
c_\pm X_\pm \xi =1$. Thus, we should fix the parameters $c_\pm$ as
$c_\pm=1/(X_\pm \xi)$, with $X_\pm$ as given in \reef{X+} and
\reef{X-} respectively.

Continuity of the $g_{\vf\vf}$ metric component imposes the
following  relation between the curvatures in the three bulk
regions and the parameters $c_i$ \be c_0 R_0 =c_\pm R_\pm ~.
\label{cont2} \ee

Since the gauge field strength is continuous through the
codimension-1 brane  and taking into account the above relation,
the solution for the gauge field which vanishes at the poles
remains exactly the same as before \reef{Anorth}, \reef{Asouth}.
Thus the quantization condition \reef{N} is the same as in the
codimension-2 model.

The solution of the Goldstone fields depends as discussed in the
previous section, on the periodicity of $\vf$ and the
normalization of $\s_\pm$ in the original Higgs theory. With the
assumption we made in the previous section (\ie  $C=1/\xi$ with
$C$ the normalization of $\s_\pm$) the Goldstone fields are simply
\be \s_\pm=n_\pm \vf ~~~~,~~~~ {\rm with}~~~n_\pm \in \mathbb{Z}~.
\ee

The junction conditions will determine the brane parameters
$\la_\pm$, $v_\pm$ and they will also give  a relation between the
quantum numbers of the brane scalar field $n_\pm$ and the bulk
gauge field $N$. In Appendix B we present all the necessary steps
to arrive at the extrinsic curvatures
$\hat{K}_{\hat{\m}\hat{\n}}$. Using these quantities we can
compute the  junction condition \reef{Kjunction}. Its  $(\vf \vf)$
component reads \be
 \pm 4 M^4 \left.{z' \over z} \sqrt{f}\right|_{r_\pm} \left({1 \over R_0}-{1 \over R_\pm}\right) = \la_\pm - {v_\pm^2 (n_\pm-E {\cal A}_\vf^{\pm})^2 \over 2  c_0^2 R_{0}^2  f(r_\pm)} ~, \label{ffeqn}
\ee
and its $(\m\n)$ component is given by
\be
\pm  \left. M^4 \sqrt{f} \left(  3  { z' \over z}   + {1 \over 2}{ f' \over  f  } \right)  \right|_{r_\pm}\left({1 \over R_0}-{1 \over R_{\pm}}\right)
=\la_{\pm} + {v_{\pm}^2 (n_{\pm}-E {\cal A}_\vf^{\pm})^2 \over 2 c_0^2 R_{0}^2  f(r_\pm)} ~. \label{mneqn}
\ee
Furthermore, the gauge field junction condition \reef{Fjunction} can be easily evaluated as
\be
\mp   c_0 R_0 M^2  S  \left.{\sqrt{f} \over z^4}\right|_{r_\pm}  \left({1 \over R_0}-{1 \over R_{\pm}}\right) =E v_{\pm}^2 (n_{\pm}-E {\cal A}_\vf^{\pm}) ~. \label{gaugeeqn}
\ee

A relation between $n_\pm$ and $N$ can be found by taking the
difference of \reef{mneqn} and \reef{ffeqn} and dividing by
\reef{gaugeeqn}. Then, after substitution of ${\cal A}_\vf$, $c_0$
from \reef{Anorth}, \reef{Asouth},  \reef{N} and taking into
account that $E=\xi e$, we obtain 
\be 
n_\pm=\pm {N \over 2}
w_\pm(\a) ~~~ {\rm with} ~~~ w_\pm(\a)=\a^{2(1\mp 1)}{X_\pm \over
Y S} ~.\label{quant1} 
\ee\ 
It can be easily seen that it is always
$n_+-n_-=N$ because of the identity $X_+ + \a^4 X_- = 2 Y S$. In
the limit $\a \to 1$ we have $w_\pm \to 1$, and thus $n_\pm = \pm
N/2$ in agreement with \cite{PST}. It is worth noticing here that
the above relations are independent of the parameter $\xi$.

Since the quantities $n_\pm$, $N$ are integers, the above relation imposes
 a restriction to the values of the admissible warpings $\a$.  Simplifying 
 $w_+$ as
\be
w_+(\a)={2 \over (1-\a^3)}\left[{5(1-\a^8) \over 8 (1-\a^5)} -\a^3\right]~,
\ee
shows that $w_+$ is bounded as $1<w_+(\a)<5/4$. Hence, we find that for $N>0$ the scalar quantum number is restricted as $N/ 2 < n_+  < 5N/ 8$. 
    [The constraint coming from $n_-$ is complementary due
     to the relation $n_+-n_-=N$]. This excludes warped solutions
      for small monopole numbers  $N \leq 4$. The first warped
       solution exists for $N=5$, with $n_+=3$, $n_-=-2$
        and for warping $\a \approx .44$.

The above restriction dictates that not all warped solutions can
be regularized  in the way we have described in the previous
section. A possible way to understand why, is that, due to the 
compactness of the space, there are topological constraints for the 
various fields (bulk gauge field and brane scalar field) which may not 
be possible to be satisfied simultaneously for any warping $\a$. A similar 
conclusion would be reached if we had used a brane  3-form field 
$B_{\hat{\m}\hat{\n}\hat{\k}}$, 
instead of the scalar field $\s$, with a coupling $B_{(3)} \wedge A_{(1)}$,
 instead of $\s A_{(1)}$. However, the similarity of the conclusion 
would be due 
to the duality of the two dynamical systems in five dimensions. We are not
 aware of any other 
brane action which generates a brane energy momentum tensor of the anisotropic 
type that we should have for the particular regularization to work. Thus, we 
cannot conclude whether the above restriction is due to some 
fundamental physical obstruction, or it is an artifact of the specific 
regularization procedure which simply lowers the codimension.  The later conjecture, however, is conceivable and it would certainly be 
interesting to investigate alternative brane actions which would clarify 
the generality of our result.

From the difference of \reef{mneqn} and \reef{ffeqn} and dividing
by the square of \reef{gaugeeqn}, we can  obtain an expression for
the parameters $v_\pm$ as \be E^2 v_{\pm}^2= \pm  \left.{3(1-\a)
\sqrt{f} \over 2 z^4 \left( {1 \over z^3}-{5 (1-\a^8) \over 8 \a^3
(1-\a^5)} \right) }\right|_{r_\pm}  \left( {1 \over R_0} -{1 \over
R_{\pm}} \right)~, \ee which for the limit $\a \to 1$ reduces to
$e^2 v_{\pm}^2= \pm {\sqrt{1-r_{\pm}^2} \over r_{\pm} c_0 R_0}
(1-c_0)$, in agreement with \cite{PST}. In the  codimension-2
limit $r_\pm \to \pm 1$, the  parameters $v_\pm$ vanish regardless
of the warping.

Finally, taking the sum of \reef{mneqn} and \reef{ffeqn} we obtain
the values of the brane tensions
 \be
 \la_{\pm}=\pm \left. {M^4
\over 20 (1-\a)\sqrt{f}}\left( -8z +{11 \over 2}\cdot {1-\a^8
\over 1-\a^3}\cdot{1 \over z^4 }-4\a^3{1-\a^5 \over 1-\a^3}\cdot{1
\over z^7} \right)\right|_{r_{\pm}}  \left( {1 \over R_0} -{1
\over R_{\pm}} \right) \label{lasol}
 \ee
 which  for the limit $\a
\to 1$ reduces to $\la_{\pm}=\pm {M^4 r_{\pm} \over 2 c_0
R_0\sqrt{1-r_{\pm}^2}}(1-c_0)$, in agreement with \cite{PST}. In
the  codimension-2 limit $r_\pm \to \pm 1$, the  brane tensions
$\la_\pm$ diverge.

 The latter divergence is not worrisome when comparing with the
codimension-2 limit,  because the  quantity with physical meaning 
 is the total tension  
of the ring. This  is 
obtained by integrating the four-dimensional part of the brane 
energy momentum tensor $t^{(\pm)}_{\m\n}$ over the azimuthal 
direction and reads
\be T_{\pm}= \int d\vf  \sqrt{g_{\vf
\vf}} ~t^{(\pm)}_{00}=
 2\pi \xi c_0 R_{0}  \sqrt{f(r_{\pm})}\left[ \la_{\pm} + {v_{\pm}^2 (n_{\pm}-E {\cal A}_\vf^{\pm})^2 \over 2 c_0^2  R_{0}^2  f(r_{\pm})}\right]~.
\ee 
The divergent part of $\la_\pm$ is canceled by $\sqrt{f(r_{\pm})}$.  
Then, evaluating the bracket from \reef{mneqn}, we can take the
codimension-2 limit $r_\pm \to \pm 1$. The tensions in this limit
read \be T_\pm=2\pi M^4\left(1-c_0 X_\pm \xi \right) \ee and
coincide with the tensions  of the codimension-2 case presented in
section \ref{co2back}.

\section{Supersymmetric warped compactifications}

In this section we will study  the supersymmetrized version of the
previous model. To do so, we consider the gauge supergravity model
of Salam and Sezgin \cite{SS} where gravity is coupled to  a
 six-dimensional gauge field ${\cal A}_M$, a Kalb-Ramond field ${\cal B}_{MN}$
and  a dilaton field $\x$  in a way that respects ${\cal N}=2$
supersymmetry. The general axisymmetric solutions of such a model
involve compact spaces with sphere topology and two codimension-2
singularities at the poles of the deformed sphere
\cite{Gibbons,burgess2,burgess1}. The bosonic action of the system
(neglecting the Kalb-Ramond field which can be consistently set to
zero background value)  is given by 
\be
 S= \int d^6x \sqrt{-g} \left( {M^4 \over 2} R  -{1 \over
4}e^{\x/M^2}{\cal F}^2 -{1 \over 2}(\de_M \x)^2 -4g_0^2M^4
e^{-\x/M^2} \right) - T_\pm\int d^4x \sqrt{-\g_{\pm}} ~,~~~  \label{ssaction}
\ee
where   $T_\pm$ are the tension  of the codimension-2 branes
situated at the north  and south pole respectively and $g_0$ is
the gauge coupling  of the gauged $U(1)_R$ of the model. The
metric $\g^{\pm}_{\m\n}$ is the induced metric on the two 3-branes
and $M$ the six-dimensional Planck mass.

In the absence of the brane terms, \ie without codimension-2 
singularities, 
the vacuum of solution of  \cite{SS} respects ${\cal N}=1$ supersymmetry. 
However, the brane terms in \reef{ssaction}, necessary for warping to be 
allowed,  break supersymmetry explicitly to ${\cal N}=0$. In that respect 
we are not strict by
calling the compactification supersymmetric. Nevertheless, we will keep this 
terminology since the solution stems from a gauged supergravity action with 
 the addition of only localized terms which break supersymmetry.

The equations of motion in the bulk are easily obtained as
\bea
R_{MN}&=& 2g_0^2 e^{-\x/M^2}g_{MN} + {1 \over  M^4} \de_M \x \de_N \x \nn \\
&+& {1 \over  M^4}e^{\x/M^2}\left[{\cal F}_{MK}{\cal F}^{~K}_N - {1 \over 8}{\cal F}^2 g_{MN}\right]
~,~~~~~~\label{Einsteins} \\
\square^{(6)} \x &=& {1 \over  4 M^2}e^{\x/M^2}{\cal F}^2-4g_0^2M^2e^{-\x/M^2} ~,  \label{Scalars} \\
\de_M \left( \sqrt{-g}~e^{\x/M^2} {\cal F}^{MN}\right)&=&0 ~.
\label{Gauges} \eea 

The above equations of motion have the scaling  symmetry 
\be
g_{MN} \to u~ g_{MN}~~~~~,~~~~~\x \to \x + M^2 \log u ~. \label{scale} 
\ee 
Note that this is not a symmetry of the action
since $S \to u^2 S$.  By applying this scaling symmetry to one of
the  solutions of the equations of motion, we can obtain a new
(and inequivalent) solution.

In the following, we will first discuss the background solution
with codimension-2 branes  and then we will present the procedure to
regularize these branes by lowering their codimension, in the same way we
accomplished it in the non-supersymmetric case.

\subsection{The general solution with codimension-2 branes} \label{co2backsusy}

The general bulk axisymmetric solution of the above system, with a
monopole in the internal space was provided in  \cite{Gibbons,burgess2}.
 The metric has, as
in the non-supersymmetric case, a black hole form in the extra two
dimensions. The warped solution reads explicitly \bea
ds_6^2&=& \r ~\eta_{\m\n}dx^\m dx^\n  + {dr^2 \over F} + c_0^2 F d\th^2 ~,\label{metricbhsmis}\\
{\cal F}_{\r\th}&=& -{b_0 \over \r^3}~, \label{metricbhs}\\
\x &=& M^2 \log \r ~, \label{scalarbhs} \\
{\rm with} ~~~ M^4 F(\r)&=&-2g_0^2M^4 \r -{b_0^2 \over 4 c_0^2 \r^3}+{\m_0
\over \r}~. \label{gaugebhs}
\eea

Observe that the above solution has the same structure as the
non-supersymmetric solution (\ref{metricbhmis})-(\ref{gaugebh})
with the difference that $\rho$ appears here with a lower power.
This appears to be crucial in the following.

As before, we will consider the space $0<\r_-<\r<\r_+$ between two
roots of $F(\r)$. The parameters $b_0$ and $\m_0$ can be expressed
as a function of $\r_+$ and $\a \equiv \r_- / \r_+$ as \bea
b_0&=&\r_+^2 {M^2 c_0 \a \over R_0}~, \\
\m_0&=& \r_+^2 {M^4 (1+\a^2) \over 4 R_0^2} ~.
\eea

 To make the comparison with the unwarped case more
transparent, we will make the coordinate transformation \bea
\r=\r_+ z ~~ {\rm with} ~~ z(r)={1 \over 2}[(1-\a)r+(1+\a)] ~, \\
\vf= {(1-\a) \over 2 R_0^2} \th ~, \eea with $R_0^2=1/(8g_0^2)$
representing the average radius of the internal space.  In the new
coordinates the solution \reef{metricbhsmis}-\reef{gaugebhs}
becomes \bea
ds_6^2&=& \r_+ \left\{ z \eta_{\m\n} dx^\m dx^\n + R_0^2\left[ {dr^2 \over f} +c_0^2 f d\vf^2 \right] \right\} ~, \label{metricbhz}\\
{\cal F}_{r \vf}&=& -c_0 R_0M^2 S \cdot {1 \over z^3}~, \label{gaugebhz} \\
\x &=& M^2 \log (\r_+ z)~, \label{scalarbhz} \\
{\rm with} ~~~ f&=& {4  R_0^2 \over \r_+(1-\a)^2}F ~, \\
{\rm and} ~~~ S&=&\a ~.
\eea
From the above definition we can verify that $f(r)$ depends only on the parameter $\a$ as
\be
f(r)= {1 \over  (1-\a)^2}\left[ -z
-{\a^2 \over z^3 }+{(1+\a^2) \over z }\right]~.
\ee

For the range of the coordinate $\vf$ we make the same assumption
as before, \ie there is  $\vf \in [0,2\pi \xi)$, with $\xi$ a
function of $\a$ with the limit that as $\a \to 1$ it is $\xi(\a)
\to 1$. The two codimension-2 branes are situated at $\r=\r_\pm$,
or in the new coordinates at $r(\r_\pm)=\pm 1$ with $z(\r_+)=1$,
$z(\r_-)=\a$. Expanding the function $f$ at $r \to \pm 1$ we get
$f \to 2 (1\mp r) X_\pm$ with the quantities $X_\pm$ given by \be
X_+= {1+\a \over 2} ~~~~~,~~~~~ X_-= {1+\a \over 2 \a^2} ~.
\label{Xs} \ee

Then the metric around  $r=\pm 1$ reads
\be
ds_6^2 \approx  \r_+ \left\{ z(\pm 1)  \eta_{\m\n} dx^\m dx^\n + {R_0^2 \over X_\pm} \left[ {dr^2 \over 2 (1\mp r)} + c_0^2 X_\pm^2  2 (1\mp r) d \vf^2 \right] \right\} ~. \label{localconicals}
\ee

The deficit angles at the two singularities are thus $\d_\pm=2\pi (1-\b_\pm)$
 with $\b_\pm=c_0 X_\pm \xi$ and are supported by codimension-2 branes with tensions
  $T_{\pm}=M^4 \d_\pm$. For completeness, we present in Appendix C
   the relation of the above gauge with the one used in \cite{Gibbons}.

The scaling symmetry \reef{scale} is manifested in the solution by
the appearance  of the quantity $\r_+$.  For the rest of the
paper, we will choose $\r_+=1$. We  stress, however, that this is
{\it not} a gauge choice as in the non-supersymmetric case, but a
mere choice of a subset of solutions (from which all solutions can
be obtained by applying the symmetry).

It is straightforward to compute the flux quantization condition
of the gauge field. The gauge field in the northern and southern
patches of the manifold are given as before  by \bea
{\cal A}^+_\vf&=&  c_0 R_0 M^2 {S \over (1-\a)}\left({1 \over z^2}-1 \right)\label{Anorths}~,\\
{\cal A}^-_\vf&=&  c_0 R_0 M^2 {S \over (1-\a)} \left({1 \over
z^2}-{1 \over \a^2} \right)\label{Asouths}~, \eea which in the $\a
\to 1$ limit gives the correct limit \reef{gaugeunwarped}.  Then
single valuedness of the gauge transformation at the overlapping
region, gives the quantization condition \bea
2 c_0 R_0 M^2 \xi e ~Y  &=& N   ~~~,~~~N \in \mathbb{Z}~,  \label{Ns}\\
{\rm with}~~~Y&=&{1+\a \over 2 \a^2 }S~.
\eea

A choice $\xi=2\a/(1+\a)$ is particularly helpful, since then
the quantization condition remains the same as in the case of no
warping and the deficit angles are given by $\b_+=c_0 \a$ and
$\b_-=c_0/ \a$.

\subsection{The regularization}

We will regularize the model in the same way that we did it for
the non-supersymmetric case, by the introduction of codimension-1
branes. Using similar notation as before, the action of the system
is written as 
\bea 
S= \int_{{\cal M}_i} d^6x \sqrt{-g} \left({M^4
\over 2} R  -{1 \over 4}e^{\x/M^2}{\cal F}^2 -{1 \over 2}(\de_M
\x)^2 -4g_i^2M^4 e^{-\x/M^2} \right) \nonumber
\\ - \int_{\S_\pm} d^5x \sqrt{-\g_\pm}\left(V_\pm(\x) +{v_\pm^2
\over 2} (\tilde{D}_{\hat{\m}} \s_\pm)^2 +M^4 \{{\cal K}\}_\pm\right)~,
\eea 
where $g_i$,~$i=0,\pm$ are the gauge couplings in the
different bulk regions  and  $V_\pm(\x)$ are the potentials for
the dilaton on the  codimension-1 branes. The internal space has
the shape given in  Fig.~\ref{internalfig}. The equations of
motion in the bulk are the same as \reef{Einsteins}-\reef{Gauges}
and $g_0$ is substituted by $g_i$ for the three bulk regions. The
junction conditions on the branes are obtained by matching the
surface terms from each side of the brane with the variation of
the brane action and read 
\bea
\{\hat{K}_{\hat{\m}\hat{\n}}\}_\pm&=&-{1 \over M^4}\left[-V_\pm \g_{\hat{\m}\hat{\n}}+ v_\pm^2 \left( \tilde{D}_{\hat{\m}} \s_\pm \tilde{D}_{\hat{\n}} \s_\pm -{1 \over 2} ( \tilde{D}_{\hat{\k}} \s_\pm)^2 \g_{\hat{\m}\hat{\n}} \right)\right] ~, \label{Kjunctions}\\
\{e^{\x/M^2}n_{M} {\cal F}^{M}_{~N}\de_{\hat{\k}} X^N  \}_\pm
&=&-E v_\pm^2 \tilde{D}_{\hat{\k}} \s_\pm ~, \label{Fjunctions}\\
\{n_{M} \de^{M} \x \}_\pm&=&{d V_\pm \over d \x } ~. \label{chijunctions}
\eea

\subsection{The general solution with codimension-1 branes}

The background solution of the above regularized system will be
presented and we will show that taking the limit of $r_\pm \to \pm
1$, the solution of Section \ref{co2backsusy} with the
codimension-2 branes is correctly reproduced.

The solution in the bulk region ${\cal M}_0$ between the two
branes ($r_-<r<r_+$) (see Fig.~\ref{internalfig})
 is the one described in Section \ref{co2backsusy} \bea
ds_6^2&=&  z~ \eta_{\m\n} dx^\m dx^\n + R_0^2\left[ {dr^2 \over f} +c_0^2 f d\vf^2 \right] ~, \\
{\cal F}_{r \vf}&=& -c_0 R_0 M^2  S \cdot {1 \over z^3}~, \\
\x &=& M^2 \log  z~, \eea with $R_0^2=1/(8g_0^2)$. On the other
hand the caps ${\cal M}_\pm$,  with $r_+<r<1$ and $-1<r<r_-$
respectively,  are described by the solutions \bea
ds_6^2&=&  z~ \eta_{\m\n} dx^\m dx^\n + R_\pm^2\left[ {dr^2 \over f} +c_\pm^2 f d\vf^2 \right] ~, \\
{\cal F}_{r \vf}&=& - c_\pm R_\pm M^2  S \cdot {1 \over z^3}\\
\x &=& M^2 \log  z  ~, \eea with $R_\pm^2=M^4/(2\La_\pm)$. In
order that there is no deficit angle in the caps (\ie when
$\b_\pm= c_\pm X_\pm \xi =1$), we should demand that
$c_\pm=1/(X_\pm \xi)$, with $X_\pm$ as given in \reef{Xs}.

Continuity of the $g_{\vf\vf}$ metric component imposes again  the
relation \be c_0 R_0 = c_\pm R_\pm ~. \label{cont2s} \ee Since the
gauge field strength is continuous through the codimension-1 brane
and taking into account the above relation, the solution for the
gauge field which vanishes at the poles remains exactly the same
as before \reef{Anorths}, \reef{Asouths}. Thus the quantization
condition \reef{Ns} is the same as in the codimension-2 model.

The solution of the Goldstone fields, with the normalization
discussed  in Appendix A, is given as before 
\be 
\s_\pm=n_\pm \vf
~~~~,~~~~ {\rm with}~~~n_\pm \in \mathbb{Z}~. 
\ee 
The brane
parameters $v_\pm$ and the value and the slope of the dilaton
potential  $V(\x)$ (which implicitly depends on branes positions)
will be provided by the junction conditions. Furthermore, we will
obtain a relation between the quantum numbers of the brane scalar
field $n_\pm$ and the bulk gauge field $N$. In Appendix B, we
present all the necessary steps to arrive at the extrinsic
curvatures $\hat{K}_{\hat{\m}\hat{\n}}$. Using these extrinsic
curvatures, the $(\vf \vf)$ component of the junction condition
\reef{Kjunctions} reads \be
 \pm 2 M^4 \left.{z' \over z} \sqrt{f}\right|_{r_\pm} \left({1 \over R_0}-{1 \over R_\pm}\right) = V_\pm - {v_\pm^2 (n_\pm-E {\cal A}_\vf^{\pm})^2 \over 2c_0^2 R_{0}^2  f(r_\pm)} \label{ffeqns} ~,
\ee
and its $(\m\n)$ component is given by
\be
\pm  \left. {M^4 \over 2} \sqrt{f} \left(  3  { z' \over z}   + { f' \over  f  } \right)  \right|_{r_\pm}\left({1 \over R_0}-{1 \over R_{\pm}}\right)
=V_{\pm} + {v_{\pm}^2 (n_{\pm}-E {\cal A}_\vf^{\pm})^2 \over 2 c_0^2 R_{0}^2  f(r_\pm)} ~.  \label{mneqns}
\ee
Furthermore, the gauge field junction condition \reef{Fjunctions} can be easily evaluated as
\be
\mp  c_0  R_0 M^2 ~ S \left.{\sqrt{f} \over z^2}\right|_{r_\pm}  \left({1 \over R_0}-{1 \over R_{\pm}}\right) =E v_{\pm}^2 (n_{\pm}-E {\cal A}_\vf^{\pm})~,  \label{gaugeeqns}
\ee
and the scalar field junction condition \reef{chijunctions} gives
\be
\mp M^2 \left.{z' \over z} \sqrt{f}\right|_{r_\pm} \left({1 \over R_0}-{1 \over R_\pm}\right)= {d V_\pm \over d \x } ~. \label{scaleqns}
\ee
In the above equations we have denoted $V_\pm=V(\x(r_\pm))$ and  $d V_\pm / d \x = (d V / d \x) (\x(r_\pm))$.

A relation between $n_\pm$ and $N$ can be found by taking the
difference of \reef{mneqns}  and \reef{ffeqns} and dividing by
\reef{gaugeeqns}.  Then, after substitution of ${\cal A}_\vf$,
$c_0$  from \reef{Anorths}, \reef{Asouths}, \reef{Ns} and taking
into account that $E=\xi e$, we obtain the simple result \be
n_\pm= \pm {N \over 2}~. \label{quant2}\ee From the above relation
we see that the warping $\a$ is {\it not} restricted as in the
non-supersymmetric case. Furthermore, the Goldstone field quantum
number is related to the gauge field quantum number in exactly the
same way as in the unwarped case. However, as we noted in the 
non-supersymmetric case, we cannot be sure if this conclusion depends on 
the particular regularization, before we try some alternative brane 
action.

From the difference of \reef{mneqns} and \reef{ffeqns} and
dividing by the square  of \reef{gaugeeqns}, we can  obtain an
expression for  the parameters $v_\pm$ as \be E^2 v_{\pm}^2= \pm
\left.{2 \a^2(1-\a) \sqrt{f} \over   ( 2\a^2- (1+\a^2)z^2)
}\right|_{r_\pm}  \left( {1 \over R_0} -{1 \over R_{\pm}}
\right)~, \ee which for the limit $\a \to 1$ reduces to $e^2
v_{\pm}^2= \pm {\sqrt{1-r_{\pm}^2} \over r_{\pm} c_0 R_0}
(1-c_0)$. In the  codimension-2 limit $r_\pm \to \pm 1$, the
parameters $v_\pm$ vanish regardless of the warping.

Taking the sum of \reef{mneqns} and \reef{ffeqns} we obtain the
values  of the dilaton potential on the rings \be V_{\pm}=\pm
\left. {M^4 (1-\a)  \over 4} ~\sqrt{f}~{(4z^4-3(1+\a^2)z^2+2\a^2)
\over z (\a^2-z^2)(1-z^2)}\right|_{r_{\pm}}  \left( {1 \over R_0}
-{1 \over R_{\pm}} \right) ~, \label{Vsols} \ee which  for the
limit $\a \to 1$ reduces to $V_{\pm}=\pm {M^4 r_{\pm} \over 2 c_0
R_0\sqrt{1-r_{\pm}^2}}(1-c_0)$. In the  codimension-2 limit $r_\pm
\to \pm 1$, the  above values  $V_\pm$ diverge. On the other hand
\reef{scaleqns} provides the values of the slope of the dilaton
potential on the rings. In the limit  $\a \to 1$ these slopes
vanish.

 As we had explained in the non-supersymmetric case, the divergence 
of $V_\pm$ in the
codimension-2 limit is not worrisome. The total
tension of the ring is obtained by integrating the
four-dimensional part of the brane energy momentum tensor
$t^{(\pm)}_{\m\n}$ over the azimuthal direction. This is given by
\be 
T_{\pm}=\int d\vf \sqrt{g_{\vf \vf}} ~t^{(\pm)}_{00}= 2\pi \xi
c_0 R_{0} \sqrt{f(r_{\pm})}\left[ V_{\pm} + {v_{\pm}^2 (n_{\pm}-E
{\cal A}_\vf^{\pm})^2 \over 2 c_0^2 R_{0}^2 f(r_{\pm})}\right]~.
\ee
 Then, evaluating the bracket in the above relation using
\reef{mneqns} and taking the codimension-2 limit $r_\pm \to \pm
1$, we can find the tensions in this limit  
\be
 T_\pm=2\pi
M^4\left(1-c_0 X_\pm \xi \right)~, 
\ee
 which coincide with the
tensions of the codimension-2 case presented in section
\ref{co2backsusy}.

\section{Conclusions}

In the present work, we discussed the regularization of the
codimension-2  singularities in Einstein-Maxwell systems with  a
monopole in the internal space. We have applied the regularization
procedure of \cite{PST}, which lowers the codimension of the
branes, in axisymmetric backgrounds with general warping. In this
way, instead of working with codimension-2 branes, which have
rather restrictive gravitational dynamics, one can work with
ring-like codimension-1 branes which are much better understood.

An important ingredient of the  regularization was the
Goldstone-like  dynamics of a brane  scalar field coupled to the
bulk gauge field. We have shown that the above procedure works for
any warped solution in the supersymmetric case, when the quantum
number of the brane scalar field and the bulk gauge field are
related as in (\ref{quant2}). On the contrary, in the
non-supersymmetric case, not all warpings are consistent with the
corresponding quantum number constraint (\ref{quant1}). This may,
however, be an artifact of the scalar field dynamics which have
been used to generate the necessary energy momentum tensor needed
for the regularization to work.

An important application of the above regularizations would be the
study of cosmology on the codimension-1 branes and its limit as
the radius of these branes goes to zero (codimension-2 limit).
This will shed light on the cosmological properties of the
codimension-2 branes \cite{ClineGiova,thick}. A particular case,
is the one in which the ring-like brane moves in the static bulk
background and induces a mirage cosmological evolution on it, as
\eg in  \cite{mirage}. We plan to address this issue in a
forthcoming publication \cite{work}.

\section*{Acknowledgments}

We would like to thank L. Sorbo for helpful discussions. This work
is co-funded by the European Social Fund and National Resources
-(EPEAEK II)-PYTHAGORAS. E.P is partially supported  by the
European Union through the Marie Curie Research and Training
Network UniverseNet (MRTN-CT-2006-035863).

\def\theequation{A.\arabic{equation}}
\setcounter{equation}{0}
\vskip0.8cm
\noindent
{\Large \bf Appendix A: Origin of the Goldstone field action}
\vskip0.4cm
\noindent

In this Appendix we will discuss the origin of the brane scalar
field action and justify the comments we made in the main text.
Let us consider a complex Higgs field  $H$ coupled to the gauge
field ${\cal A}_{\hat{\m}}$.  We know that since we have a compact
dimension, every field living in it that couples to ${\cal
A}_{\hat{\m}}$, should have a charge which is an integer multiple
of a fundamental charge $e$. For convenience we will assume that
the Higgs' charge is equal to the fundamental charge $e$ (although
it could in principle be an integer multiple). The Higgs kinetic
Lagrangian  reads 
\be 
-(D_{\hat{\m}} H)(D_{\hat{\n}} H)^*
\g^{\hat{\m} \hat{\n}}~, 
\ee 
with $D_{\hat{\m}} H =\de_{\hat{\m}} H
-ie {\cal A}_{\hat{\m}}H$ the covariant derivative with respect to
the gauge field. At low enough energies we can integrate out the
massive  radial part and  focus only on the Goldstone mode of the
Higgs. The latter is given by $H={\rm v}~ e^{i \S}$ and thus the
Lagrangian becomes
 \be
-{\rm v}^2(\bar{D}_{\hat{\m}} \S)(\bar{D}_{\hat{\n}} \S)
\g^{{\hat{\m}}{\hat{\n}}}~, 
\ee 
with $\bar{D}_{\hat{\m}} \S =
\de_{\hat{\m}} \S - e {\cal A}_{\hat{\m}}$,  not to be confused as
covariant derivative but to be just  a convenient abbreviation. If
we use the angular coordinate $\vf \in [0,2\p \xi)$ as in the main
text, the solution for the field $\S$ which respects the
periodicity of the space is $\S= n \vf/\xi$ with $n \in {\mathbb
Z}$. Had we used  an  angular coordinate  $\F \in [0,2\p)$, the
solution would be $\S= n \F$. We see that the periodicity of the
space directly enters in the solution of the brane scalar field.

However, we  do not know in principle from the effective brane
Lagrangian the normalization of the Goldstone field. Let us for
example rescale the Goldstone  field as $\S=C \s$. Then the Higgs
field is written as $H={\rm v}~ e^{i C \s}$ and the Lagrangian
reads 
\be 
-{\rm v}^2 (\bar{D}_{\hat{\m}} [ C\s
])(\bar{D}_{\hat{\n}} [C\s]) \g^{{\hat{\m}}{\hat{\n}}} = -v^2
(\tilde{D}_{\hat{\m}} \s )(\tilde{D}_{\hat{\n}} \s )
\g^{{\hat{\m}}{\hat{\n}}}~, 
\ee 
with $v=C {\rm v}$,
$\tilde{D}_{\hat{\m}} \s= \de_{\hat{\m}} \s - E {\cal
A}_{\hat{\m}}$ and $E= e/C$ a redefined ``charge''. The latter
brane Lagrangian was used in the main text. The  solution for the
field $\s$  is then $\s= \S/C =  n \vf/(C\xi)$ with $n \in
{\mathbb Z}$.

In the above presentation, there are two ambiguities: the
normalization  $C$ of the Goldstone field  (\ie if it is $\s$ or
$\S$ that appears in the action) and the  parameter $\xi$ which
has to do with the range of the angular coordinate. We can fix
part of this ambiguity by imposing that $C=1/\xi$. The solution
for the field $\s$ (which respects the periodicity of the space)
is then  $\s= n \vf$ with $n \in {\mathbb Z}$, independent of the
quantity $\xi$. The Higgs vev  is then related to the parameter
$v$ appearing in the action as $v={\rm v}/\xi$ and the ``charge''
$E$ is given by $E=\xi e$. Let us note that the remaining
ambiguity (that of the determination of $\xi$) can be further
partially fixed by comparison with the unwarped solution as
described in the main text.

We should stress here  that the ``charge'' $E$ of $\s$ need not
be a integer multiple of the fundamental charge $e$. This is
because it is only the charge of the parent Higgs field  appearing
in the original Higgs Lagrangian that is subject to quantization.
The fact that there is a non-integer ``charge'' for $\s$ is merely
due to the normalization of the  Goldstone mode and the range of
the angular variable.

\def\theequation{B.\arabic{equation}}
\setcounter{equation}{0}
\vskip0.8cm
\noindent
{\Large \bf Appendix B: Extrinsic curvatures}
\vskip0.4cm
\noindent

In this Appendix we will calculate the extrinsic curvatures on the
brane positions,  which will be used in the main text to evaluate
the junction conditions. Let the brane position in the bulk be $X^M(x^{\hat{\m}})$, from which we evaluate the induced metric on the brane as $\g_{\hat{\m}\hat{\n}}=g_{MN}\de_{\hat{\m}} X^M \de_{\hat{\n}} X^N $. Firstly, we should calculate the normal vector of the brane  $n_M$, which is orthogonal to all the tangent vectors of the brane 
$\de_{\hat{\n}} X^M$, that is $\de_{\hat{\n}} X^M n_M =0$. The tangent vector in the brane time direction $u^M \propto \de_{0} X^M $ normalized to unit norm $u^M u^N g_{MN}=-1$ is known as the proper velocity of the brane. The normal vector is normalized as
 $n_M n_N g^{MN} =1$.

Once the normal vector is computed, one can
evaluate the projection  tensor $h_{MN}=g_{MN}-n_M n_N$, which is related to 
the induced metric as $h^{MN}=\g^{\hat{\m}\hat{\n}}\de_{\hat{\m}} X^M \de_{\hat{\n}} X^N$ and satisfies also $\g_{\hat{\m}\hat{\n}}=h_{MN}\de_{\hat{\m}} X^M \de_{\hat{\n}} X^N $ due to the orthogonality of the normal to the tangent vectors. Afterwards, the extrinsic curvature  is  given 
by ${\cal K}_{MN}= h^{K}_ {M} h^{\La}_ {N} \nabla_K n_\La$ (the covariant 
derivative computed with $g_{MN}$) and with trace ${\cal K} = g^{MN} {\cal K}_{MN}$. The pullback of the  extrinsic curvature on the brane is given by 
$K_{\hat{\m}\hat{\n}}={\cal K}_{MN} \de_{\hat{\m}} X^M \de_{\hat{\n}} X^N$ and has the property that $K=\g^{\hat{\m}\hat{\n}} K_{\hat{\m}\hat{\n}}={\cal K}$. The combination which appears in the junction conditions is $\hat{K}_{\hat{\m}\hat{\n}}=K_{\hat{\m}\hat{\n}}-K
\g_{\hat{\m}\hat{\n}}$.

In our examples, since the branes are static, their velocity
vectors  are $u^M \propto (1,\vec{0},0,0)$. The normal vectors of
the branes with the above mentioned normalization are given then
by $n_M=\sqrt{g_{rr}} (0,\vec{0},1,0)$. Since the $(rr)$ component
of the metric in the gauge that we work is discontinuous, we have
different normal vectors pointing {\it inwards} the  cap
$n_{M}^{in}$ and the {\it outwards} to the  bulk $n_{M}^{out}$.
Their non-vanishing components, for the upper and the  lower brane
are  respectively \be n_{r}^{in}=\pm {R_\pm \over \sqrt{f(r_\pm)}}
~~~~~,~~~~~ n_{r}^{out}=\mp {R_0 \over \sqrt{f(r_\pm)}}
~.\label{normals} \ee

The projection tensors $h_{MN}=g_{MN}-n_M n_N$ on the two branes
are then $h_{\hat{\m}\hat{\n}}=g_{\hat{\m}\hat{\n}}$ and
$h_{rr}=h_{r\hat{\m}}=0$. Let us now split the presentation for
the non-supersymmetric and the supersymmetric case.

In the {\bf non-supersymmetric case}, the  induced metrics $\g^\pm_{\hat{\m}\hat{\n}}$
on the branes are given simply by 
\be 
ds_{5(\pm)}^2= z^2(r_\pm)
\eta_{\m\n} dx^\m dx^\n + R_{\pm}^2 c_\pm^2 f(r_\pm)  d \vf^2~.
\ee 
The non-zero extrinsic curvature components then read
$K_{\m\n}=-\G^r_{\m\n}n_r$ and $K_{\vf \vf}=-\G^r_{\vf\vf}n_r$.
The six-dimensional Christoffel symbols that we need are \be
\G^r_{\m\n}=-{z z' f \over R_0^2}\e_{\m\n}~~~,~~~\G^r_{\vf\vf}=-{1
\over 2}c_0^2 ff'~, \ee with $' \equiv d/dr$. With the above, we
compute the extrinsic curvatures $\hat{K}_{\hat{\m}\hat{\n}}$.
Inwards to the caps, for the upper and the lower brane, they are
respectively \be \hat{K}^{in}_{\m\n} = \mp \left.{\eta_{\m\n}
\over R_\pm }  z^2 \sqrt{f} \left(  3  { z' \over z}   + {1 \over
2}{ f' \over  f  } \right)  \right|_{r_\pm} ~~~~~,~~~~
\hat{K}^{in}_{\vf \vf} = \mp \left.4 c_\pm^2 R_\pm {z' \over z}
f^{3/2}\right|_{r_\pm}~. \ee Outwards to the bulk, for the upper
and the lower brane, they are respectively \be
\hat{K}^{out}_{\m\n} = \pm \left. {\eta_{\m\n} \over R_0 }z^2
\sqrt{f}  \left(  3  { z' \over z}   + {1 \over 2}{ f' \over  f  }
\right)  \right|_{r_\pm} ~~~~~,~~~~~ \hat{K}^{out}_{\vf \vf} = \pm
\left.4 c_0^2 R_0 {z' \over z} f^{3/2}\right|_{r_\pm}~. \ee

In the {\bf supersymmetric case}, the   induced metrics $\g^\pm_{\hat{\m}\hat{\n}}$  on the
branes are  given by 
\be 
ds_{5(\pm)}^2= z(r_\pm) \eta_{\m\n} dx^\m
dx^\n + R_{\pm}^2 c_\pm^2 f(r_\pm)  d \vf^2~. \ee The
six-dimensional Christoffel symbols that we need are \be
\G^r_{\m\n}=-{ z' f \over 2
R_0^2}\e_{\m\n}~~~,~~~\G^r_{\vf\vf}=-{1 \over 2}c_0^2 ff'~, 
\ee
With the above, we compute the extrinsic curvatures
$\hat{K}_{\hat{\m}\hat{\n}}$. Inwards to the caps, for the upper
and the lower brane, they are respectively \be \hat{K}^{in}_{\m\n}
= \mp \left.{\eta_{\m\n} \over 2 R_\pm }  z \sqrt{f} \left(  3  {
z' \over z}   + { f' \over  f  } \right)  \right|_{r_\pm}
~~~~~,~~~~ \hat{K}^{in}_{\vf \vf} = \mp \left.2 c_\pm^2 R_\pm {z'
\over z} f^{3/2}\right|_{r_\pm}~. \ee Outwards to the bulk, for
the upper and the lower brane, they are respectively \be
\hat{K}^{out}_{\m\n} = \pm \left. {\eta_{\m\n} \over 2 R_0 }z
\sqrt{f}  \left(  3  { z' \over z}   + { f' \over  f  } \right)
\right|_{r_\pm} ~~~~~,~~~~~ \hat{K}^{out}_{\vf \vf} = \pm \left.2
c_0^2 R_0 {z' \over z} f^{3/2}\right|_{r_\pm}~. \ee

With the above it is straightforward to evaluate the junction
conditions  \reef{Kjunction}, \reef{Fjunction} in the
non-supersymmetric case and the junctions \reef{Kjunctions},
\reef{Fjunctions}, \reef{chijunctions} in the supersymmetric case.

\def\theequation{C.\arabic{equation}}
\setcounter{equation}{0}
\vskip0.8cm
\noindent
{\Large \bf Appendix C: An alternative  gauge in the supersymmetric case }
\vskip0.4cm
\noindent

In this Appendix we will present the relation of the gauge that we used for the solution in the supersymmetric case  with the one used in  \cite{Gibbons} (slightly rescaled). Let us make the coordinate transformation 
\bea 
z(r)&=&W(R)^2 ~,\\
 \psi &=& {1+\a \over 2} \vf   ~,
\eea
where the function $W(R)$ is defined as 
\be
 W(R)=\left(\frac{f_1}{f_0}\right)^{1/4}, \ \ f_0=1+\frac{R^2}{4}, \label{m2}\ \ f_1=1+\frac{R^2}{4}\a^2~.
\ee

Then it is straightforward to verify that the solution \reef{metricbhz}, \reef{gaugebhz}, \reef{scalarbhz} is transformed to the one of \cite{Gibbons}
\bea
ds_6^2&=& \r_+ \left\{ W^2 \eta_{\m\n} dx^\m dx^\n + R_0^2 L^2\left[ dR^2 +c_0^2 A^2 d\psi^2 \right] \right\}  ~, \\
{\cal F}_{R \psi}&=& M^2 c_0 R_0 \a \cdot {L^2 A \over W^6}  ~, \\
\x &=& M^2 \ln (\r_+ W^2) ~,
\eea
with
\be
L(R)={W \over f_0}~~~~{\rm and}~~~~  A(R)={ R \over W^4}~.
\ee

Note that here we keep $c_0$, although in \cite{Gibbons} a choice of $c_0=1$ was made. The range of the angular coordinate has been altered by this transformation, as $\psi \in \left[\left. 0,2\pi \xi {1+\a \over 2} \right)\right.$. To compare with \cite{Gibbons}, where $\psi \in [0,2\pi)$, one should make the choice $\xi = 2 / (1+\a)$. Then the deficit angles are given by $\b_+=c_0$ and $\b_-=c_0/\a^2$. Note that in the conventions of \cite{Gibbons} there is $\a=q/(4g)$, with $q$ the magnetic charge and $g$ the gauge coupling.

\end{document}